\documentclass[aps,twocolumn,floats,floatfix,superscriptaddress]{revtex4-1}
\usepackage{graphicx,amssymb,amsmath}
\usepackage[utf8]{inputenc}
\usepackage{color}

\definecolor{red}{rgb}{1,0,0}
\definecolor{green}{rgb}{0,1,0}
\definecolor{blue}{rgb}{0,0,1}

\DeclareMathOperator{\sinc}{sinc}

\begin{document}
\title{Superextreme Waves Generation in the Linear Regime}
\author{Cristian Bonatto}
\author{Sandra D. Prado}
\affiliation{Instituto de F\'\i sica, Universidade Federal do Rio Grande do Sul, 91501-970 Porto Alegre, Brazil}
\author{Fernando L. Metz}
\affiliation{Instituto de F\'\i sica, Universidade Federal do Rio Grande do Sul, 91501-970 Porto Alegre, Brazil}
\affiliation{Departamento de F\'isica, Universidade Federal de Santa Maria, 97105-900 Santa Maria, Brazil}
\affiliation{London Mathematical Laboratory, 14 Buckingham Street, London WC2N 6DF, United Kingdom}
\author{J\'ulio R. Schoffen}
\affiliation{Instituto de F\'\i sica, Universidade Federal do Rio Grande do Sul, 91501-970 Porto Alegre, Brazil}
\author{Ricardo R. B. Correia}
\affiliation{Instituto de F\'\i sica, Universidade Federal do Rio Grande do Sul, 91501-970 Porto Alegre, Brazil}
\author{Jandir M. Hickmann}
\affiliation{Instituto de F\'\i sica, Universidade Federal do Rio Grande do Sul, 91501-970 Porto Alegre, Brazil}


\begin{abstract}

Extreme or rogue waves are large and unexpected waves appearing with higher probability than predicted by Gaussian statistics.
 Although their formation is explained by both linear and nonlinear wave propagation, nonlinearity has been considered a necessary ingredient to generate superextreme waves, i.e., an enhanced wave amplification, where the wave amplitudes exceed by far those of standard rogue waves.
 Here we show, experimentally and theoretically, that superextreme optical waves emerge in the simple case of linear one-dimensional light diffraction. The underlying physics is a long-range correlation
 on the random initial phases of the light waves.
 When subgroups of random phases appear recurrently along the spatial phase distribution, a more
 ordered phase structure greatly increases the probability of constructive
 interference to generate superextreme events, i.e., non-Gaussian statistics with super-long tails.
 Our results consist in a significant advance in the understanding of extreme waves formation by linear superposition of random waves, with
 applications in a large variety of wave systems.
\end{abstract}
\maketitle



In the last decade, a tremendous deal of attention has been devoted to the investigation, in physical systems, of extreme or rogue waves, i.e., large amplitude waves appearing more often than predicted by the normal distribution \cite{onorato}. Examples arise in hydrodynamics \cite{kharif1}, optics \cite{solli}, plasmas \cite{bailung}, acoustics \cite{ganshin}, and quantum systems \cite{bludov}.

A central topic in rogue waves studies is the understanding of the main factors that lead to deviations from the normal distribution. A deep comprehension of the mechanisms associated with the emergence of long-tail statistics
  is usefull for the more ambitious task of predicting and controling the occurrence of unexpected large events. This is of practical interest, since extreme events can be potentially destructive in different contexts, ranging from damage of maritime structures to breakdown of optical and communication systems.
  Apart from such practical motivations, there is also a fundamental interest in the phenomenon itself, since the emergence of rogue waves is a complex phenomenon which has triggered considerable efforts of an interdisciplinary  scientific community to provide a deep understanding of its generation. Generally speaking, scientists are interested in understanding how highly coherent structures can emerge from disordered and small amplitude systems, being, at same time, statistically significant.


In recent years, remarkable advances have been done in this direction, with noteworthy contributions coming from theoretical developments and controlled experiments 
with electromagnetic and water waves, in linear and nonlinear wave propagation.
In linear systems, it has been shown that when correlation or inhomogeneity is present in the wave field propagation, an increased probability of generating rogue waves is observed \cite{white,heller,hohmann,arecchi,metzger,liu,mathis}.
In nonlinear systems, a wealth of wave phenomena has been shown to be relevant for the emergence of rogue waves. In this context,   
  modulational instability has received considerable attention \cite{onorato2,kharif,janssen,solli2,dyachenko2,baronio,kibler}.
Other mechanisms in nonlinear systems include: the dynamics of partially coherent waves \cite{koussaifi}, the directional properties of the waves \cite{onorato3}, and the wind force \cite{toffoli}, in the case of water waves; weak nonlinearities without modulation instability in ocean waves \cite{fedele}; a variety of nonlinear scenarios ranging from chaotic and turbulent flows to soliton and breather collisions in optics \cite{toenger,walczak,hammani,borlaug,churkin,montina,bonatto1,selmi,walczak2,kasparian,pisarchik,lecaplain,odent,bosco,gibson,pierangeli1,pierangeli2}.

More recently, considerable attention has been directed to the investigation of the possibility to generate extremely high energy concentration, i.e., super coherent or super localized structures, from small-amplitude disordered fields. The term super rogue waves has been coined to describe new classes of waves exhibiting an enhanced wave amplification, as shown experimentally and theoretically in a water wave tank \cite{chabchoub}.
Super rogue waves correspond to higher-order solutions of the focusing nonlinear Schr\"odinger equation, a paradigmatic model for studying rogue waves phenomena,
and exhibit significant higher amplitudes that increase progressively as the order of the solution increases \cite{akhmediev2,chabchoub2}.
In optics, superextreme light pulses
have been predicted to occur in a CO$_2$ laser under harmonic modulation \cite{bonatto2}.

The current development stage of rogue waves formation suggests that linear effects can generate only an initial wave amplification, while nonlinearities are responsible for extra wave amplification or focusing, leading to much higher wave amplitudes than those observed in the purely linear case \cite{onorato2,heller,dyachenko1,matheakis,safari}. Concerning superextreme waves, the enhanced focusing behavior  has been observed only in the presence of nonlinearities.
An important open question is whether super coherent structures can be generated from a small-amplitude random field by a purely linear interference, exhibiting, at the same time, a significant probability of occurrence.

In this Letter, we provide a positive answer to the question above by demonstrating experimentally and numerically the generation of superextreme waves by the simple linear superposition of random waves in an optical system. By studying spatial memory effects of diffracted light waves, we provide the underlying physics behind superextreme waves formation in the linear regime. We show that,
when disordered phases exhibit long-range correlations in space,
the so-called linear interference model generates waves with amplitudes as high as those observed in systems with strong nonlinearities.
It is remarkable that, long after Lord Rayleigh developed his famous statistics when investigating random superposition of sound waves \cite{rayleigh} -- which forms the basis of the so-called Gaussian model for the surface elevation of water waves \cite{longuet-higgins} --
  and after the recent progresses of rogue waves formation in linear systems \cite{white,heller,hohmann,arecchi,metzger,liu,mathis},
  we can still find a new and relevant qualitative dynamical scenario in the
  linear interference model.


\begin{figure}
  \includegraphics[width=8.5cm]{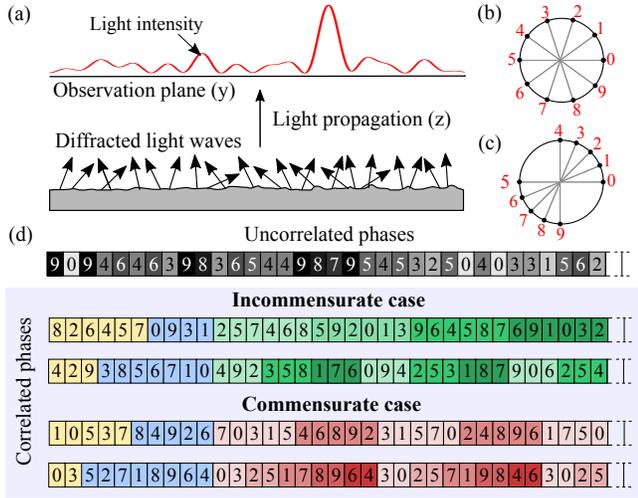}
  \caption{(a) Illustration of the one-dimensional wave diffraction. (b) and (c) Examples of distinct sets of $L=10$ initial phases, represented by the integers $0$ to $9$.
    (d) Illustration of uncorrelated and correlated initial random phases, corresponding to memoryless, $M=4$, $M=7$, $M=5$, and $M=8$, from top to bottom. Each integer corresponds to a phase, as shown in (b) or (c). See text for details.
}
\label{fig:1}
\end{figure}

We consider the simple physical situation of linear and one-dimensional light diffraction of a certain number of random waves, as illustrated in Fig.~\ref{fig:1}(a). Initially, the waves present equal amplitudes and uniformly distributed random phases, with distinct degrees of correlation.
After the waves propagate in free space, a diffraction pattern is observed in the far field (Fraunhofer plane).
We consider an even number $L$ of distinct initial phases in the interval $[0,2{\pi})$ such that the overall sum of the phasors is zero. The $L$ phases can be equally spaced in the unit circle (spanning all the four quadrants) [Fig.~\ref{fig:1}(b)] or have another phase configuration, such as $L/2$ phases in the first quadrant and $L/2$ phases in the third quadrant [Fig.~\ref{fig:1}(c)].
  Here, we choose the latter case with $L=10$ \footnote{We have chosen the initial phases distributed only in the first and third quadrants due to experimental limitations, since our spatial light modulator does not imprint phases larger than $3\pi/2$. Numerical simulations with other initial phase configurations, such as equally distributed phases along the four quadrants, have shown the same qualitative results, as we show in the supplemental material \cite{sm}}.

  \begin{figure}
  \includegraphics[width=8.5cm]{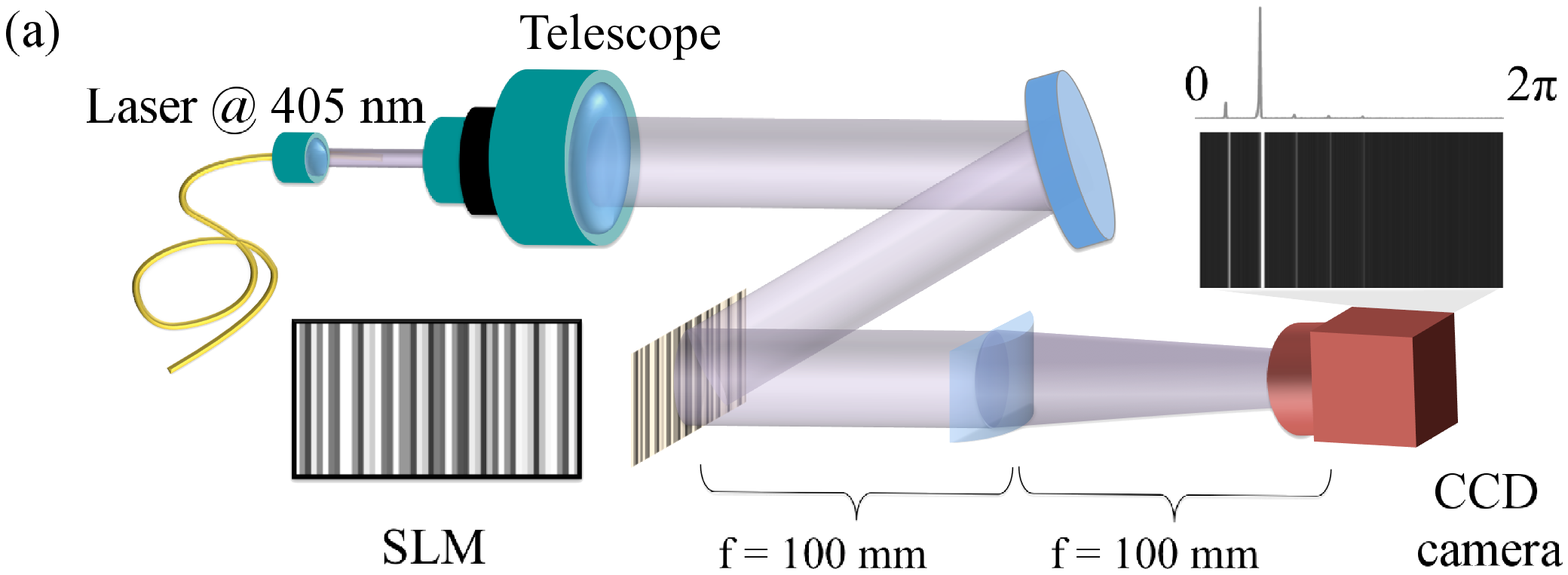}\\
  \includegraphics[width=8.5cm]{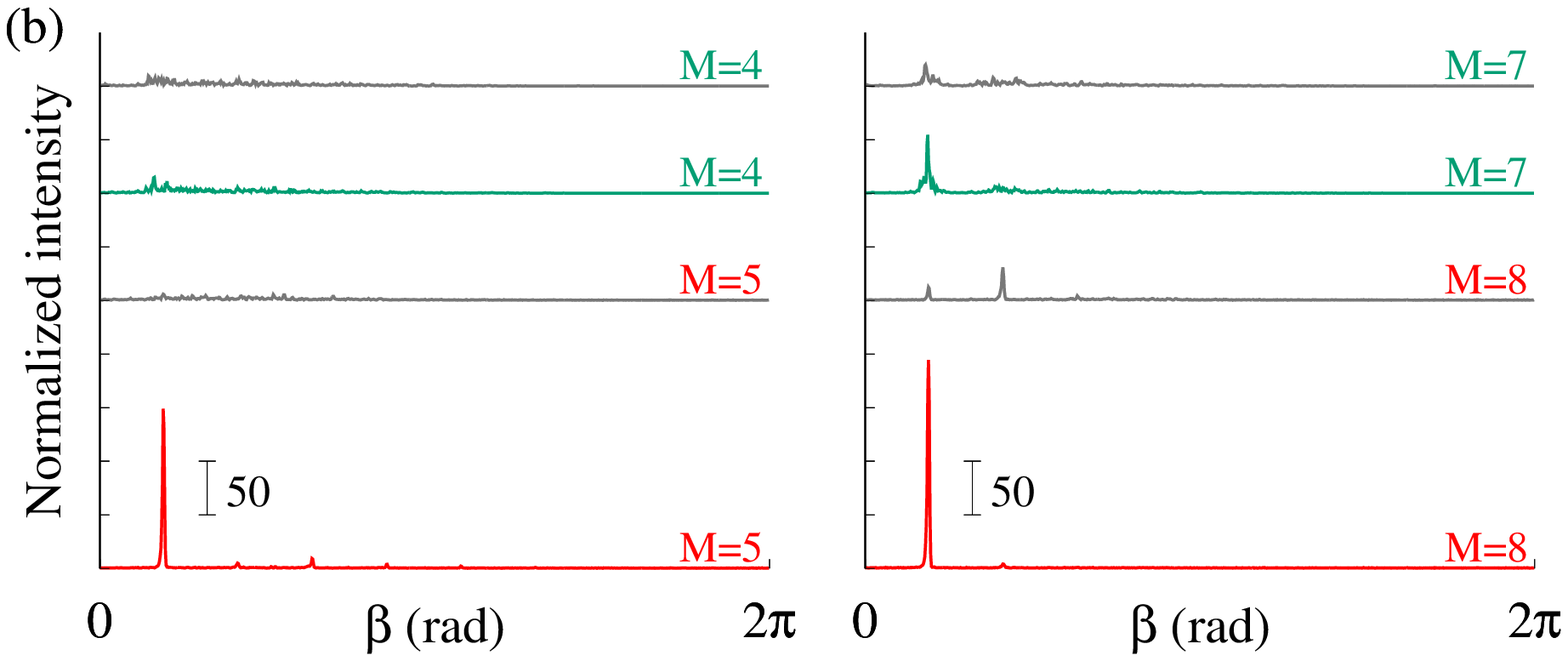}
  \caption{
    (a) Experimental setup and illustrations of random input phases and a measured diffraction pattern. 
    (b) Measured diffraction patterns showing the highest measured intensity peak for each corresponding memory lentgth, for the IC (green) and CC (red). Gray plots show the diffraction patterns containing the most frequent maximum intensity from a total number of 1000 realizations. The intensity is normalized by the average intensity of each diffraction pattern and $\beta$ is related to the diffraction angle.}
\label{fig:2}
\end{figure}

  \begin{figure*}
   \includegraphics[width=18cm]{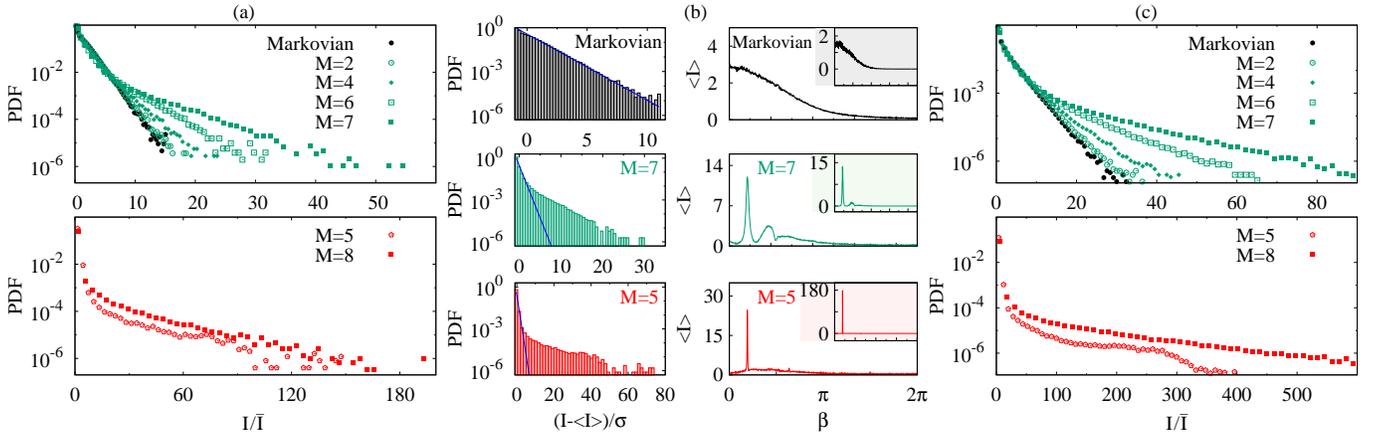}
  \caption{
    (a) Probability density function (PDF) of the measured light intensity.
    (b) Left: PDFs of the light intensity normalized by its standard deviation $\sigma$. The blue straight line denotes the negative exponential fitting. Right: Average intensity $\langle I \rangle$ and variances (insets) as a function of screen position $\beta$.
    The overall average $\overline{I}$ and the standard deviation $\sigma$ are obtained by averaging over all realizations
    and all points recorded on the camera,
    for each correlation degree. 
  (c) PDFs from numerical simulations.}
  \label{fig:3}
\end{figure*}

Without any correlation, the spatially distributed initial random phases are statistically independent and we obtain a Markovian (memoryless) phase sequence. This uncorrelated case (UC) is illustrated by the gray shade sequence in Fig.~\ref{fig:1}(d).
To generate correlated random phases, we use a variation of a two-dimensional scheme of input phases proposed to investigate diffraction patterns of non-Markovian light \cite{fischer}.
We start a correlated phase sequence by taking a random permutation of the $L$ distinct phases. The last $M$ random phases, shown in blue color in Fig.~\ref{fig:1}(d), define the spatial memory length.
Then, we perform a random permutation of the remaining $L-M$ phases [shown in yellow color in Fig.~\ref{fig:1}(d)] and place them to the right of the $M$-phase block.
We now identify the last $M$ phases of the newly-formed sequence and repeat successively the previously explained procedure until the total number $N$ of random waves is reached. In this way, a group of $L-M$ random phases always depends on the previous $M$ phases, resulting in a non-Markovian phase sequence.

The sequences of correlated phases with different memory lengths can be classified in two distinct cases, which we denominate here commensurate case (CC) and incommensurate case (IC). The CC occurs when the ratio $L/(L-M)$ is an integer. In this case, the phase sequence contains distinct subsets of phases with no common elements. For example, for $M=5$ there are two distinct subsets of five random phases, and for $M=8$ there
are five subsets of two random phases [see Fig.~\ref{fig:1}(d)].
In the CC, the spatial configuration of initial phases typically forms a quasiperiodic sequence, where the same group of random phases appears recurrently, what greatly enhances the coherence properties of the waves and favors a quasiresonant process. As we show below, this process is crucial to generate superextreme waves.

Our experiments are performed in the following way. One-dimensional random phase sequences, generated in a computer according to the procedure described above, are imprinted on the spatial light modulator (SLM), which acts as a phase mask. Since the SLM is a two-dimensional structure and we are interested in investigating the simplest case of one-dimensional wave diffraction, the one-dimensional phase sequences are repeated along the rows of the SLM [a sketch of the SLM is shown in Fig.~\ref{fig:2}(a), where each gray shade represents a phase value]. 
The SLM is illuminated by a monochromatic light from a laser beam and the diffracted wavefront  propagates in free space, passing through a cylindrical lens, with focal distance $f=100$ mm, up to the detection in a CCD camera. Images recorded with the CCD camera exhibit a one-dimensional intensity pattern, where a single diffraction pattern is calculated averaging all rows in the image of each realization [an example of a recorded image and the respective diffraction pattern is shown in Fig.~\ref{fig:2}(a)]. This process is repeated for a large number of realizations with different random input phases. The experimental setup is shown in Fig.~\ref{fig:2}(a), and additional details are described in the supplemental material \cite{sm}.

The effects of distinct memory lengths on the measured light intensities are shown in Fig.~\ref{fig:2}(b). The parameter $\beta= k a \sin{\theta}$ is related to the diffraction angle $\theta$, where $k$ is the magnitude of the wavevector of the incident wave and $a$ is the pixel width of the SLM. Figure \ref{fig:2}(b) shows the measured diffraction pattern containing the highest intensity peak for some memory lengths corresponding to the IC (green curves) and CC (red curves). We can see that the the memory lengths of the CC, $M=5$ and $M=8$, produce much higher light intensity peaks than the memory lengths of the IC, $M=4$ and $M=7$.
The gray curves in Fig.~\ref{fig:2}(b) show typical realizations of single diffraction patterns
where the highest intensity peak equals the mode of the statistical distribution of the maxima intensities.
In other words, if we take a random realization,  the highest intensity peak of
the diffraction pattern is more likely to coincide with the highest intensity in the gray curves.

In Fig.~\ref{fig:3}(a), we show, for distinct correlation degrees, a plot of the probability density function (PDF) of the intensity $I$ normalized by its overall average $\bar{I}$, obtained from all different realizations and all points from the camera. In the UC (black points), the intensity follows a negative exponential distribution, which is a signature of Rayleigh statistics. 
In this case, the high intensity events have an extremely low probability of occurrence. In the IC (green points), events with higher intensities are more likely to occur than in the UC. As memory increases, deviations from Rayleigh statistics become more pronounced, leading to long-tailed PDFs. This is the usual signature of extreme waves. In the CC (red points), the PDFs exhibit superlong tails, due to what we call superextreme waves. Below we discuss the quantitative and qualitative differences between extreme and superextreme waves. Note that superextreme light intensity events, such as those observed for $M=5$ and $M=8$, are very unlikely to occur in the IC, even when considering  $M=7$, a case of high memory length.

A usual criterion to define extreme events is the comparison of the intensity (or amplitude) of an event with the average intensity (or amplitude) plus a certain number of standard deviations \cite{dysthe,bonatto1}. Here, instead of providing an arbitrary definition for extreme and superextreme events, we plot the PDF of $I-\langle I \rangle$ normalized by its standard deviation $\sigma$ [see Fig.~\ref{fig:3}(b)]. Thus, the horizontal scales of the PDF graphs in Fig.~\ref{fig:3}(b) indicate directly the number of standard deviations by which the measured intensity exceeds the average intensity. The intensity distributions in the CC exhibit much stronger deviations from Rayleigh statistics than in the IC. In Fig.~\ref{fig:3}(b) we also show the average and the variance of the light intensity along the screen, obtained from averaging over all diffraction patterns.

A simple linear model reproduces qualitatively the experimental results. We model the SLM as a one-dimensional array with a total number of $N$ pixels, each one characterized by a finite width and a random phase. An incident monochromatic wave diffracts on the SLM, propagates in free space, and yields a resultant electric field on the Fraunhofer plane
\begin{equation}
  E=E_0\sinc{\left(\beta/2\right)}\sum\limits_{m=1}^{N} e^{i m\beta+i \phi_m},\\
  \label{eq:1}
\end{equation}
where $E_0$ is the amplitude of the incident electric field 
and $\phi_m$ is the random phase of pixel $m$. For the theoretically computed diffraction patterns, we performed numerically the summation in Eq.~(\ref{eq:1}) by using $E_0=1$ and $N=1024$ random waves with the same initial phases used in the experiments and explained in Fig.~\ref{fig:1}. The statistical distributions of 1000 realizations for the numerically computed intensity $I=|E|^2$ are shown in Fig.~\ref{fig:3}(c), for distinct phase correlations. The qualitative agreement between the theoretical results of Fig.~\ref{fig:3}(c) and the experimental results of Fig.~\ref{fig:3}(a) is very good. In the supplemental material \cite{sm}, we show additional data obtained from the experiments and numerical simulations.

\begin{figure}
  \includegraphics[width=8.5cm]{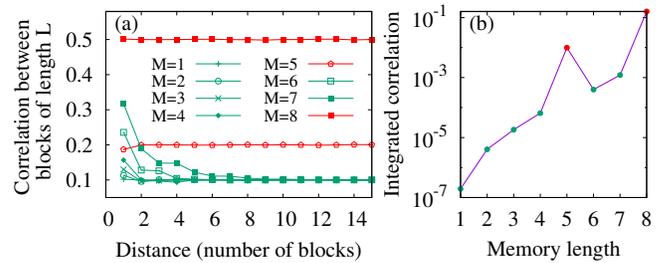}
  \caption{
    (a) Correlation between spatially separated pairs of blocks containing $L=10$ random phases as a function of their distance, measured in number of blocks along the phase sequence. The red and green curves denote the CC and IC, respectively. (b) Integrated correlation as a function of memory length. Each integrated correlation is proportional to the area under the corresponding curve shown in (a).}
\label{fig:4}
\end{figure}

The main ingredient behind the distinct qualitative behaviors of the UC, IC, and CC is the correlation properties due to memory effects of the initial phases. In order to better understand the role of the memory length in producing coherence, we numerically calculate the following correlation between two spatially separated blocks containing $L$ random phases
\begin{equation}
  C_d = \frac{1}{L}\sum_{i=1}^L  \langle \phi_i  \phi_{i + d L} \rangle,
  \label{eq:2}
\end{equation} 
where $d = 1,2,3,\dots$ is the distance between the first block with $L$ phases to the other subsequent blocks along the phase sequence. In Eq.~(\ref{eq:2}) the product $\phi_i  \phi_{i + d L}$ is equal to one if $\phi_i = \phi_{i + d L}$ and zero otherwise. The angle brackets $\langle \dots  \rangle$ denote the average over $1000$ distinct realizations of the random phases. A plot of the correlation as a function of the distance $d$ between the blocks, for fixed $L=10$ and different memory lengths, is shown in  Fig.~\ref{fig:4}(a). In the IC, the correlation decays as the distance between the blocks increases, converging to $1/L$. This means that, for large distances, the probability of finding two coherent phases is $1/L$, which is the same value as in the UC. In contrast, the correlation does not decay in the CC, having the constant value $1/(L-M)$. In other words, the probability of finding two coherent phases separated by an arbitrary number of blocks of size $L$ is $0.2$ and $0.5$ for $M=5$ and $M=8$, respectively. This is a significant difference when compared with the IC.

The overall coherence of the phase sequence, for each $M$, is obtained by computing the integrated correlation
\begin{equation}
\overline{C} = \frac{1}{n} \sum_{d=1}^{n}  \left( C_{d}- \frac{1}{L} \right)^2,
\end{equation}
where $n$ is the maximum distance used in the calculation.
Since each correlation in the IC quickly decays to the value $1/L$, we used $n=50$, which already provides the asymptotic
value of $\overline{C}$.
For the CC, $C_{d}$ is independent of $d$, thus the integrated correlation is independent of $n$.
As shown in Fig.~\ref{fig:4}(b), the integrated correlation puts in a clear ordering the coherence generated by distinct memory lengths, having $M=5$ and $M=8$ as the most coherent cases. This ordering of the coherence properties of the initial phases directly reflects in the amplitudes of the extreme events generated by wave diffraction. We stress that only a large memory length is not enough for producing strong coherence. As can be seen in Fig.~\ref{fig:4}(b), a quasi-resonant memory length can generate more coherence than a longer nonresonant one.
In conclusion, by studying systematically memory effects of spatially distributed random phases of a light field, we have shown that a simple linear superposition of random waves can generate waves with superextreme amplitudes, with a significant probability of occurrence. Up to date, superextreme events have been observed, in both hydrodynamical and optical systems, only when strong nonlinearities are present. The superextreme waves are not only quantitatively, but also qualitatively distinct when compared with standard rogue waves, since they are formed by long-range phase correlations. Although we use a particular procedure for generating correlations between the optical phases of the interfering waves, other kind of long-range correlations should yield similar effects. The solutions of the linear interference model can be classified in distinct qualitative groups as a function of the phase correlations in the input waves. In two opposite limits, we have the fully disordered case (uncorrelated phases) and the fully ordered case (completely coherent phases). Random uncorrelated phases lead to the well-known Rayleigh statistics, where high-amplitude waves are very unlikely. Completely coherent phases, as found in elementary physics textbooks, lead to a {\it sinc}-like pattern. Between these two limits, we can have two distinct qualitative scenarios: short-range and long-range correlated random phases.
Short-range correlated phases lead to an increased probability of generating extreme waves, what is the usual investigated case in the rogue waves literature.
On the other hand, long-range correlations of the random phases greatly increase the probability of generating waves with superextreme amplitudes, yielding a much more pronounced L-shaped statistics when compared with standard rogue waves. 

C.B. thanks National Council for Scientific and Technological Development --- CNPq for financial support under contract number 487057/2013-9. F. L. M. thanks London Mathematical Laboratory and CNPq (Edital Universal 406116/2016-4) for financial support.

\bibliographystyle{apsrev4-1}
\bibliography{biblio.bib}

\end{document}